\documentclass[aps,prl,showpacs,preprint]{revtex4}

\usepackage{graphicx} 
\usepackage{amsmath}
\usepackage{stmaryrd}
\usepackage{MnSymbol}
\usepackage{color}

\begin{document}

\title{Generation and detection of axions using guided structures}
\author{D. D. Yavuz and  S. Inbar}
\affiliation{Department of Physics, 1150 University Avenue,
University of Wisconsin at Madison, Madison, WI, 53706}
\date{\today}
\begin{abstract}

We propose a new experimental technique to generate and detect axions in the lab with a good experimental sensitivity over a broad axion mass range. The scheme relies on using laser-based four-wave mixing, which is mediated by the hypothetical axion field. Intense pump and Stokes laser beams that are confined to a waveguide (i.e., for example, an optical fiber) with appropriately chosen frequencies resonantly drive axion generation. Under such a geometry, we predict the existence of guided axion waves, which we refer to as ``axitons". These are solutions of the axion Klein-Gordon field equation that are spatially guided by the profiles of the driving pump and Stokes laser beams. These guided axitons can then couple to a nearby fiber and mix with another laser, affecting the propagation of a probe laser beam. A key advantage of the scheme is that the mass range of the hypothetical axion can be scanned by varying the frequencies of the pump and the Stokes laser beams.  We predict that, using reasonable parameters, the technique will be able to detect axions in the mass range $10^{-6} $~eV $< m<$ $10^{-2} $~eV  with a sensitivity at the level of $10^{-12}$~GeV$^{-1}$ for the axion-photon coupling constant. 

\end{abstract} 

\maketitle

\section{Introduction and Background}

Since their first prediction about four decades ago, the interest in axions has been continually growing \cite{review,amoreview,casper,irastorza}. Light axions or axion-like-particles with a mass in the $10^{-6} $~eV $< m<$ $10^{-2} $~eV range form a compelling candidate for the dark matter in the universe \cite{darkmatter1,darkmatter2,darkmatter3}. The existence of axions would also solve one of the longstanding theoretical problems in the standard model of particle physics; the so-called strong CP problem \cite{wilczek,pq1,pq2}. Not surprisingly, there has been a large number of experimental efforts to detect this elusive particle. In this paper, we will be focusing on experimental detection schemes that rely on the interaction of the axions with electromagnetic waves. The theoretical proposals exploring axion-photon coupling date back to 80's \cite{sikivie,krauss,maiani}, spurring much experimental work. ADMX (Axion Dark Matter eXperiment) and related experiments  \cite{admx1,admx2,admx3,admx4,admx5} rely on converting low-energy axions that form the background dark matter to microwave photons using an intense DC magnetic field. The detection sensitivity of the microwave photons can be enhanced by many orders of magnitude using microwave cavities with a high quality factor (high-$Q$). Solar helioscopes such as CAST (Cern Axion Solar Telescope) \cite{cast1,cast2} and IAXO (International AXion Observatory) \cite{iaxo} use a similar mechanism, but instead search for the conversion of high energy solar axions into x-ray photons. The strategy in these experiments is to search for a change in the x-ray flux as the detector points to and away from the sun, or as earth rotates around the sun. Other ideas to detect axionic dark matter include using possible resonance effects in superconducting Josephson junctions \cite{josephson1,josephson2}. Axion dark matter detection using antiferromagnetic topological insulators has also been discussed \cite{topological}. 

The experiments mentioned above aim to detect axions that are naturally present in the environment. Another set of experiments work towards generating and detecting axions in the lab, and have greater control of their experimental parameters since they do not rely on an external source of axions. One idea is to produce axions using the interaction of an intense laser inside a Fabry-Perot cavity with a DC magnetic field, and detect the generated axions at a different location by back-converting them into photons. This set of experiments are cordially referred to as {\it light shining through a wall (LSW)}, and their sensitivity has been steadily increasing over the last few decades \cite{lsw1,lsw_new}. Other experiments along similar lines have aimed to observe a change in the phase, intensity, or the polarization of a laser, as the beam propagates through a region of space and generates axions through its interaction with a strong DC magnetic field \cite{lsw2,lsw_new2}. Currently, these laboratory-based experiments do not have detection sensitivities comparable to microwave cavity experiments. Detecting axion-like particles using interaction of multiple high-peak-power pulsed laser beams in vacuum has also been discussed \cite{mendonca,dobrich,homma1,homma2,nobuhiro}. 

In this letter, we suggest a laser-based scheme that belongs to this second group of experiments (i.e., generating and detecting axions in the lab), but also has quite high sensitivity over a broad axion mass range. As we discuss below, the scheme relies on using laser-based four-wave mixing, which is mediated by the hypothetical axion field. Intense pump and Stokes laser beams with appropriately chosen frequencies resonantly drive axion generation. The central idea of our technique is that with the pump and Stokes lasers confined to a waveguide, for example an optical fiber, the generated axions can also be guided. The spatial profiles of the driving laser beams confine the axion generation, producing propagating axion modes, which we refer to as axitons.  The physics of confinement for axitons is similar to electromagnetic modes of a fiber; instead of the radial refractive index profile, the confinement relies on the electric and magnetic beam profiles of the driving pump and Stokes lasers. These confined axitons can then couple and mix with another laser (which we term as the mixing beam), affecting the propagation of a probe laser. In the spirit of LSW experiments, the detection of the axion using mixing and probe lasers can be accomplished in a separate fiber. We predict that our scheme will be able to detect axions with a sensitivity at the level of $10^{-12}$~GeV$^{-1}$ for the axion-photon coupling constant over an axion mass range $10^{-6} $~eV $< m<$ $10^{-2} $~eV. 

Compared to traditional LSW experiments, there are four unique advantages of our scheme: (1) The mass range of the hypothetical axions can be scanned by appropriately tuning the frequencies of the pump and the Stokes laser beams. (2) Because we utilize guided modes of a fiber, the interaction length can be as much as hundreds of kilometers, which is much longer than what can be achieved using other approaches. (3) We point out the use of near-zero refractive index for the probe laser beam, which may possibly be achieved using an engineered (metamaterial) waveguide, to further enhance the detection sensitivity \cite{caspani,liberal}.  This is in the spirit of wire-metamaterials in recently suggested tunable plasma haloscopes \cite{lawson}. (4) The scheme is purely laser based and does not rely on interaction using a Tesla-level DC magnetic field. These four advantages may prove to be critical in generating and detecting this elusive particle in the near future. 
  
\section{Formalism}

We proceed with a detailed description of our scheme. Were it to exist, a scalar axion field $\phi (\vec{r},t)$  would interact with electric and magnetic fields of electromagnetism through the Lagrangian \cite{gasperini,visinelli}:
\begin{eqnarray}
\Delta \mathcal{L} = \frac{g_{a \gamma \gamma}} {\mu_0c} \phi \vec{E}  \cdot \vec{B} \quad , 
\end{eqnarray}

\noindent where the quantity $g_{a \gamma \gamma}$ is the axion-EM coupling constant. The factor $1/(\mu_0 c)$ assures that the product $g_{a \gamma \gamma} \phi$ is dimensionless.  In order to make the physics more explicit, we will use mks units in our analysis (except when we present the bounds on the coupling constant  $g_{a \gamma \gamma} $  in unit of GeV$^{-1}$, which is the tradition in the discipline). With the Lagrangian of Eq.~(1), the modified Maxwell's equations in a medium with current density $\vec{J}$ can be found through the Euler-Lagrange equations and they are:
\begin{eqnarray}
\nabla \cdot \vec{E} & = & - g_{a \gamma \gamma} c \nabla \phi \cdot \vec{B} \quad , \nonumber \\
\nabla \cdot \vec{B} & = &  0 \quad , \nonumber \\
\nabla \times \vec{E} & = & -\frac{ \partial \vec{B}}{\partial t}  \quad , \nonumber \\
\nabla \times \vec{B} & = &  \frac{1}{c^2} \frac{ \partial \vec{E}}{\partial t}  + \mu_0 \vec{J} + \frac{g_{a \gamma \gamma}}{c} \left( \frac{\partial \phi}{\partial t} \vec{B} + \nabla \phi \times \vec{E} \right) \quad . 
\end{eqnarray}

\noindent The interaction with the axion field produces additional charge and current densities, and modifies the Maxwell's equations for $\nabla \cdot \vec{E}$ and $\nabla \times \vec{B}$, while leaving the other two equations unchanged. As a result of the electromagnetic interaction, the Klein-Gordon equation for the axion field  $\phi (\vec{r},t)$  is also modified with a driving term that involves the dot product of the electric and magnetic fields:
\begin{eqnarray}
 \nabla^2 \phi - \frac{1}{c^2} \frac{ \partial^2 \phi}{\partial t^2}- \left( \frac{m c} {\hbar} \right)^2  \phi =  \frac{g_{a \gamma \gamma}} {\mu_0 c} \vec{E} \cdot \vec{B} \quad .
\end{eqnarray}

\noindent Here $m$ is the mass of the hypothetical axion and $\nabla^2$ is the Laplacian operator. The $ \vec{E} \cdot \vec{B} $ driving term produces an effective second-order nonlinearity for the vacuum and in our scheme we borrow many ideas from nonlinear optics, specifically from four-wave mixing processes \cite{boyd}.

\begin{figure}[tbh]
\vspace{-0cm}
\begin{center}
\includegraphics[width=15cm]{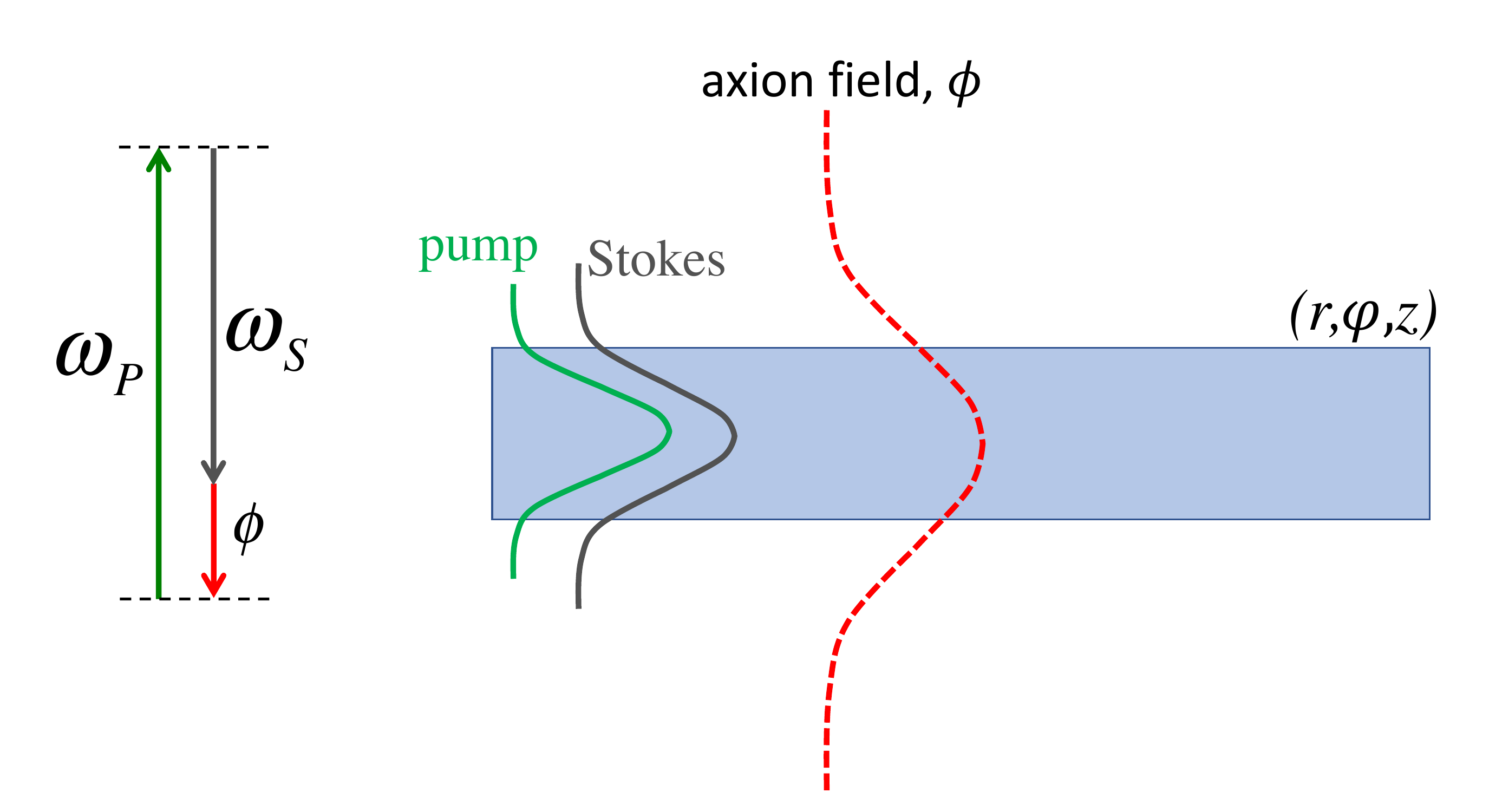}
\vspace{-0.5cm} 
\caption{\label{eit_scheme} \small   Energy level diagram and simplified schematic for producing guided axitons. Pump and Stokes laser beams whose frequency difference, $\omega_P - \omega_S$, is tuned close to the rest mass energy of the axion, resonantly drive axion generation. The two lasers are confined to an optical fiber and the solid curves are cartoon schematics for the radial profiles of the two lasers. The spatial profiles of the two beams then confine axion generation, producing guided axitons, which are shown in dashed curve.}
\end{center}
\vspace{-0cm}
\end{figure}

\section{Guided axion waves: axitons}

We first discuss the generation of axions that are confined by the spatial profiles of the pump and Stokes laser beams through the Klein-Gordon equation of Eq.~(3). Figure~1 shows the relevant energy level diagram and cartoon schematic for producing guided axion waves, which we refer to as axitons. We consider pump and Stokes laser beams propagating through a wave-guide, for example, an optical fiber. Through the $ \vec{E} \cdot \vec{B} $ term, the electric field of the pump  and the magnetic field of the Stokes will drive the axion excitation. We take the fiber to be cylindrically symmetric and assume the pump and Stokes lasers to be in specific modes of the fiber. We work in cylindrical coordinates $(r, \varphi, z)$, and take the two lasers to be of the form:
\begin{eqnarray}
\vec{E}_P (r, \varphi, z) & = & E_P u_P(r) \exp{ \left(i l_P \varphi \right)} \exp{ \left( i \beta_P z - i \omega_P t \right) } \hat{e} + c. c.  \quad , \nonumber \\
\vec{B}_S (r, \varphi, z) & = & B_S u_S(r) \exp{ \left(i l_S \varphi \right)} \exp{ \left( i \beta_S z - i \omega_S t \right) } \hat{e} + c. c.  \quad . 
\end{eqnarray}

\noindent Here $c. c.$ refers to the complex conjugate of the whole expression that is written before. $E_P$ and $B_S$ are the electric and magnetic field amplitudes for the pump and Stokes laser beams, and $\hat{e}$ denotes the common polarization direction for the two vectors (which is any direction orthogonal to the propagation direction $\hat{z}$). The quantities $u_P(r)$ and $u_S(r)$ are the radial mode functions of the corresponding lasers, and the integers $l_P$ and $l_S$ are typically referred to as orbital angular momentum numbers for the associated photons \cite{boyd}. $\beta_P$ and $\beta_S$ are the propagation constants of the modes along the fiber $z$ axis (i. e., the longitudinal $k$-vector) and the quantities $\omega_P$ and $\omega_S$ are the angular frequencies. Noting the energy level diagram of Fig.~1, the frequency difference of the pump and Stokes laser beams, $\omega_P-\omega_S$ is tuned close to the rest-mass energy of the hypothetical axion. By tuning this frequency difference, axions at different masses can be searched for. In general, there is another contribution to the generation of the axion wave, which is driven by the magnetic field of the pump beam and the electric field of the Stokes laser. For plane-wave like lasers propagating inside a bulk material, this second contribution (which is proportional to $B_P E_S^*$), would interfere destructively with the main contribution that we consider below (which is proportional to $E_P B_S^*$), reducing the produced axion amplitude. Here, we take that the relevant modes of the fiber are chosen appropriately so that this second contribution can be ignored compared to the first one. For example, one could use a TE (transverse electric) mode for the pump, whereas a TM (transverse magnetic) mode for the Stokes laser. This way, the vectors $E_P$ (electric field of the pump) and $B_S$ (magnetic field of the Stokes) can be aligned, maximizing the dot product. However, an angle would be present between $B_P$ and $E_S$, which can be tuned minimizing their dot product using specific choice of the modes.  

Because the axion generation is driven by the  $\vec{E} \cdot \vec{B} $ term, we are interested in the axion field solutions of the form:
\begin{eqnarray}
\phi (r, \varphi, z)  =  u_{\phi} (r) \exp{ \left[i (l_P -l_S) \varphi \right]} \exp{ \left[ i (\beta_P- \beta_S) z - i (\omega_P-\omega_S) t \right] } + c. c.  \quad . 
\end{eqnarray}

\noindent If the ansatz solution of the form Eq.~(5) exists, it would indicate confined axions propagating without any change in their profile along the fiber. Using Eqs.~(4) and (5), the Klein-Gordon equation for the axion field can be reduced to a single ordinary differential equation for the radial axion mode function $u_{\phi}(r)$:
\begin{eqnarray}
\frac{d^2 u_{\phi}}{d r^2} + \frac{1}{r} \frac{d u_{\phi}}{d r} - \frac{(l_P - l_S)^2}{r^2} u_{\phi}  + \left[ \frac{(\omega_P - \omega_S)^2} {c^2} - (\beta_P - \beta_S)^2 - \left(\frac{m c }{\hbar} \right)^2 \right] u_{\phi} = \frac{g_{a \gamma \gamma}} {\mu_0c} E_P B_S^* u_P(r) u_S^*(r) \quad . \nonumber \\
\end{eqnarray}

\noindent We next define the quantity $ \Delta k ^2 \equiv \frac{(\omega_P - \omega_S)^2} {c^2} - (\beta_P - \beta_S)^2 - \left(\frac{m c }{\hbar} \right)^2$, which can be thought of as an ``energy detuning": i.e., the detuning (difference) between the pump and Stokes laser energy difference (i.e., $\omega_P -\omega_S$) and the total energy (kinetic  + rest mass) of the excited axions. To elucidate the physics, we normalize the above equation through the definitions $\tilde{r} \equiv \kappa_{axion} r $  and $ \tilde{ \Delta k } \equiv \Delta k / \kappa_{axion}$. Here, the quantity $\kappa_{axion}$ is the $k$-vector of the axion particles if their kinetic energy equaled their rest-mass energy, $\kappa_{axion} = m c / \hbar$. With these definitions, Eq.~(6) reads:
\begin{eqnarray}
\frac{d^2 u_{\phi}}{d \tilde{r}^2} + \frac{1}{\tilde{r}} \frac{d u_{\phi}}{d \tilde{r}} - \frac{(l_P - l_S)^2}{\tilde{r}^2} u_{\phi}  + \tilde{ \Delta k}^2 u_{\phi} = \frac{1}{\kappa_{axion}^2} \frac{g_{a \gamma \gamma}} {\mu_0c} E_P B_S^* u_P(\tilde{r}) u_S^*(\tilde{r}) \quad . 
\end{eqnarray}

\noindent Given the parameters of the system, Eq.~(7) can now be numerically integrated to find the radial profile of the axion excitation, $u_{\phi} (\tilde{r})$. We look for physical solutions where the axion field is a maximum at $ \tilde{r} =0$, which gives the following initial condition for the first derivative:  $ \frac{d u_{\phi}}{d \tilde{r}} (\tilde{r} = 0 ) =0$. We then numerically integrate Eq.~(7), with a trial initial condition $ u_{\phi} (\tilde{r} = 0) $. For a given initial condition, the integration will typically not result in a bounded axion field, i.e., $u_{\phi}  (\tilde{r} \rightarrow \infty ) \neq 0 $, which is not physical. Given the parameters for the system, we vary the initial condition $ u_{\phi} (\tilde{r} = 0) $ until a bounded solution with $u_{\phi}  (\tilde{r} \rightarrow \infty ) = 0$ is found. 

\begin{figure}[tbh]
\vspace{-0cm}
\begin{center}
\includegraphics[width=15cm]{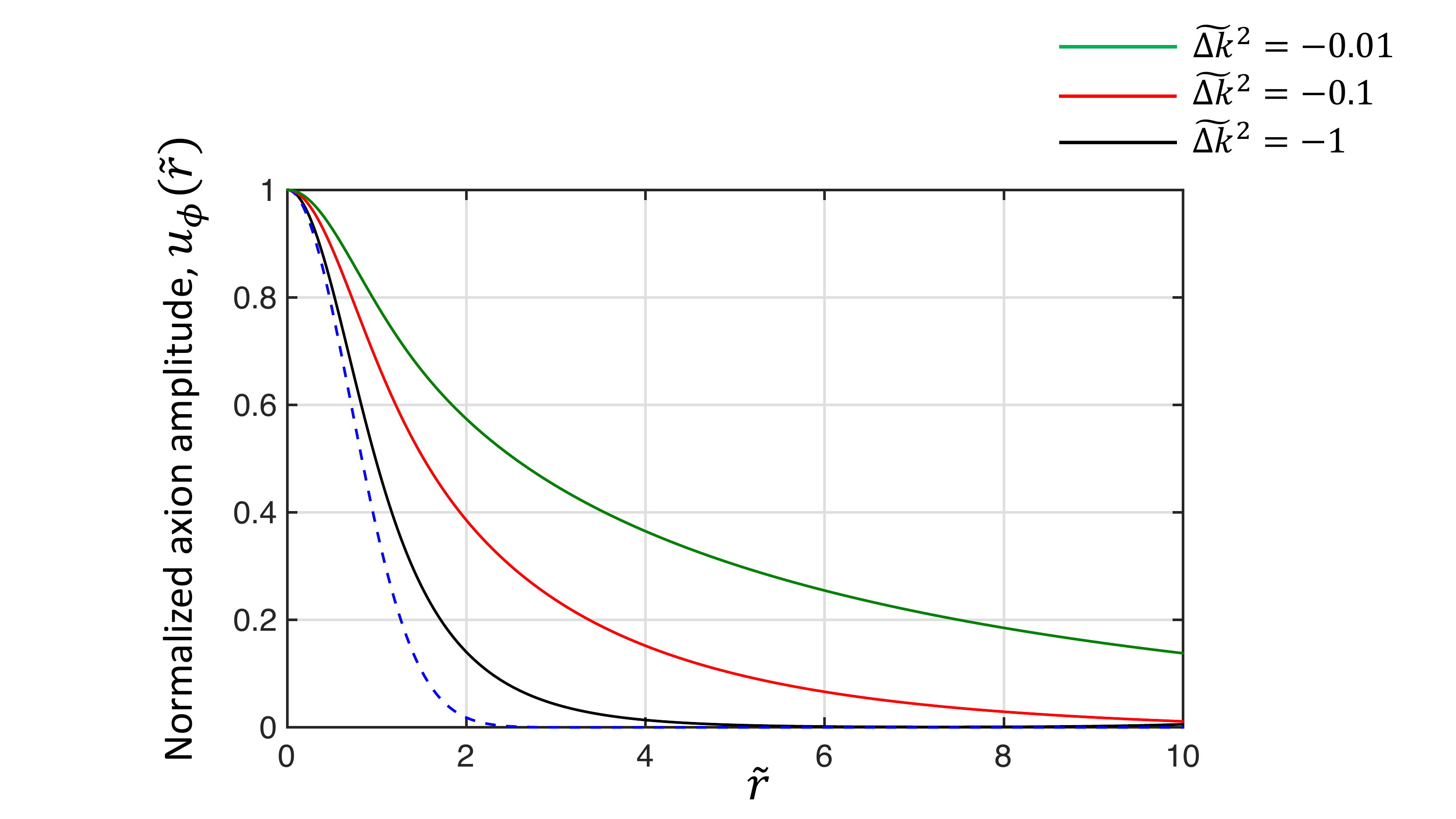}
\vspace{-0.5cm} 
\caption{\label{eit_scheme} \small   Numerically calculated normalized axiton profiles, $u_{\phi}  (\tilde{r})$, for $ \tilde{ \Delta k }^ 2 = -1 $ (solid black),   $ \tilde{ \Delta k }^ 2 = -0.1 $ (solid red), and $ \tilde{ \Delta k }^ 2 = -0.01 $ (solid green), respectively. For comparison, the dashed blue line shows the mode profiles for the driving laser beams, $ u_P (\tilde{r}) $ and $ u_S (\tilde{r}) $. For a bounded physical solution such that $u_{\phi}  (\tilde{r} \rightarrow \infty ) = 0$, we require  $ \tilde{ \Delta k }^ 2 <0 $. As the quantity  $ \tilde{ \Delta k }^ 2 $ gets closer to 0, the axion mode profile gets broader and extends significantly beyond the confinement of the driving lasers.  }
\end{center}
\vspace{-0.5cm}
\end{figure}

Figure 2 shows numerically calculated normalized axiton profiles, $u_{\phi}  (\tilde{r} )$, for $ \tilde{ \Delta k }^ 2 = -1 $ (solid black),   $ \tilde{ \Delta k }^ 2 = -0.1 $ (solid red), and $ \tilde{ \Delta k }^ 2 = -0.01 $ (solid green), respectively. For comparison, the dashed blue line shows the mode profiles for the driving laser beams, $ u_P (\tilde{r}) $ and $ u_S (\tilde{r})  $. Here, for simplicity, we take the profiles for the pump and Stokes laser beams to be Gaussian with unity width, $ u_P (\tilde{r}) = u_S^* (\tilde{r}) = \exp{(-\tilde{r}^2})$. We also take the angular momentum numbers for the pump and Stokes fields to be the same, $l_P=l_S$.  Due to well-known Bessel function solutions to differential equations of the from Eq.~(7), for a physical bounded solution such that $u_{\phi}  (\tilde{r} \rightarrow \infty ) = 0$, we require  $ \tilde{ \Delta k }^ 2 <0 $. As the quantity  $ \tilde{ \Delta k }^ 2 $ gets closer to 0, the axiton radial mode profile gets broader and extends significantly beyond the confinement of the driving lasers. This is well-illustrated in the solid green curve in Fig.~2. The quantity $ \tilde{ \Delta k }^ 2$  will likely be important parameter to tune in future experiments, since the broader the axion profile, the more it will leak to the second detection fiber as we will discuss in the detection scheme below. The numerically found initial amplitudes for the axitons, each multiplied by the scaling factor in the right hand side of Eq.~(7) (i.e., $\frac{1}{\kappa_{axion}^2} \frac{g_{a \gamma \gamma}} {\mu_0c} E_P B_S^*$) are, $ u_{\phi} (\tilde{r} = 0)= 0.23$ (for  $ \tilde{ \Delta k }^ 2 = -1 $), $ u_{\phi} (\tilde{r} = 0)= 0.48$ (for  $ \tilde{ \Delta k }^ 2 = -0.1 $), and $ u_{\phi} (\tilde{r} = 0)= 0.76$ (for  $ \tilde{ \Delta k }^ 2 = -0.01 $), respectively. Once the radial mode profile for the axion field is found numerically as shown in Fig.~2, the solution of Eq.~(5) gives the full description of the axiton mode. 

\section{Detection of axions using the guided-wave geometry}

We have so far focused on generation of confined axitons that propagate along the fiber. Now, we discuss the second half of the problem, namely how to detect these guided axitons and prove their existence. The vision that we have for a future experiment is shown in Fig. 3. Inspired by the LSW experiments, we use  a separate fiber, which we refer to as the detection fiber. This is necessary, since any material will posses a four-wave mixing nonlinearity, which would completely overwhelm the four-wave mixing interaction mediated by the axion field. We, therefore, need to make sure that the four involved laser beams do not spatially overlap. The central idea in Fig.~3 is that the axiton mode produced in the generation fiber overlaps with the detection fiber, while the pump and Stokes laser beam profiles do not. If necessary, the extinction of the pump and Stokes lasers at the detection fiber can be guaranteed by putting a metal shield between the two fibers. In the detection fiber, the axion field mixes with the magnetic field of another laser that we refer to as the mixing laser. Through the axion interaction, the mixing laser then affects the propagation (both phase and intensity) of a probe laser beam. The search for the axion relies on detecting this change on the probe laser. 

\begin{figure}[tbh]
\vspace{-0cm}
\begin{center}
\includegraphics[width=15cm]{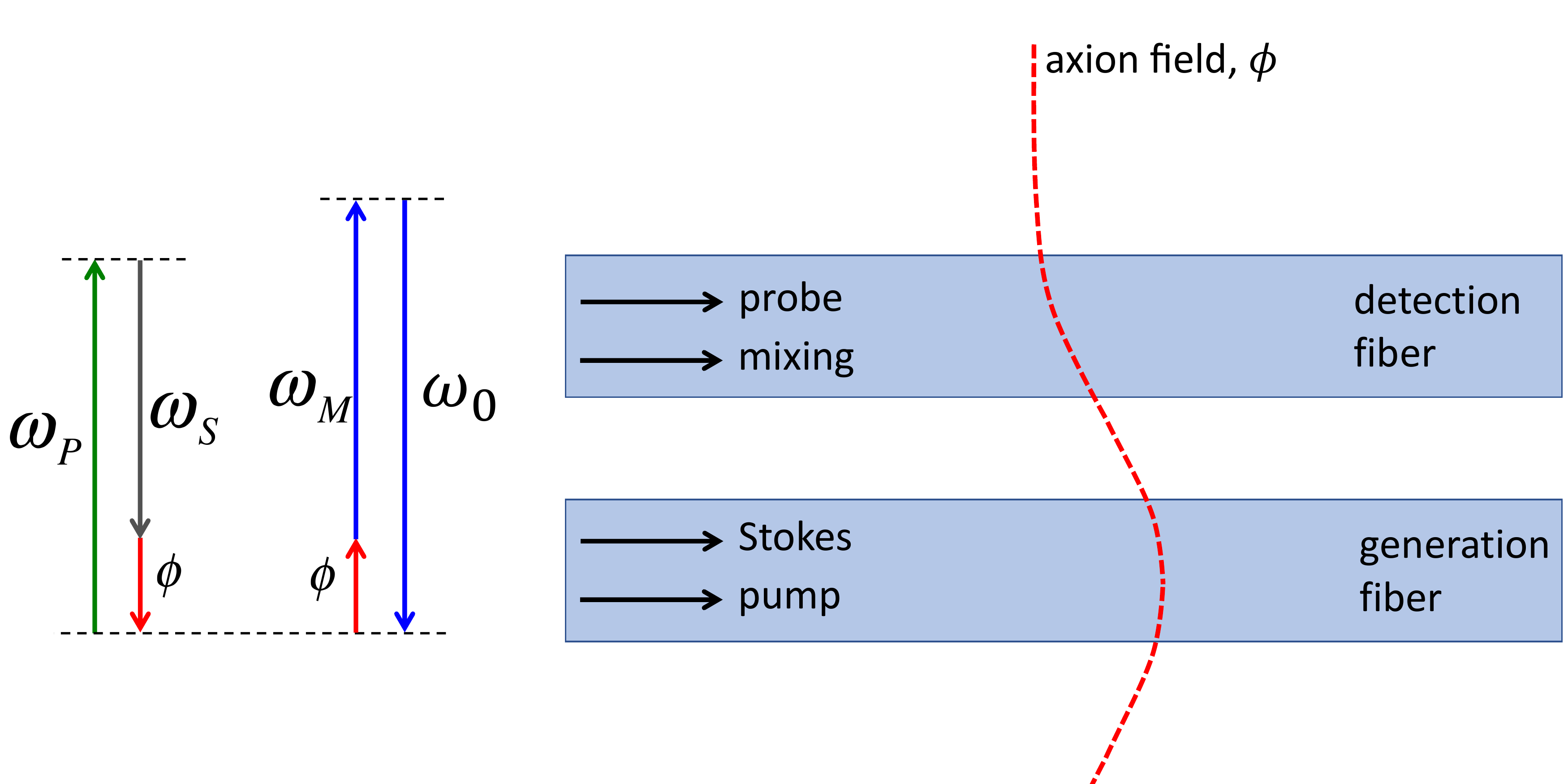}
\vspace{-0.5cm} 
\caption{\label{eit_scheme} \small  The left panel shows the energy level diagram for the four-wave mixing scheme for generating and detecting the axions. The axion field, $\phi$, produced by the pump and Stokes laser beams mix with the mixing laser, affecting the propagation of the probe laser at frequency $\omega_0$. The four-wave mixing interaction forms a closed loop: $\omega_P-\omega_S+\omega_M = \omega_0$. The mixing and the probe lasers propagate along a separate fiber, which we refer to as the detection fiber. Inspired by the LSW experiments, the axion field produced in the generation fiber (by the pump and Stokes lasers), overlaps with the detection fiber and mediates the interaction between the mixing and probe lasers.  }
\end{center}
\vspace{-0cm}
\end{figure}

Similar to above, we will now take mixing and probe lasers to be modes of the detection fiber and assume the following forms for the two waves:
\begin{eqnarray}
\vec{E}_{probe} (r, \varphi, z) & = & E_0(z) u_0(r) \exp{ \left(i l_0 \varphi \right)} \exp{ \left( i \beta_0 z - i \omega_0 t \right) } \hat{e} + c. c.  \quad , \nonumber \\
\vec{B}_M (r, \varphi, z) & = & B_M u_M(r) \exp{ \left(i l_M \varphi \right)} \exp{ \left( i \beta_M z - i \omega_M t \right) } \hat{e} + c. c.  \quad . 
\end{eqnarray}

Here, the quantity $B_M$ is the magnetic field amplitude for the mixing laser beam. To allow for a change in the phase and the intensity of the probe wave, we have explicitly made its electric field amplitude to be a function of $z$, $E_0(z)$. Without the axion interaction, the amplitude for the probe mode would be independent of distance, i.e., $E_0(z)=E_0(z=0)$. This amplitude gets modified due to axion interaction and the basic idea of the scheme is to infer the existence of axions using this amplitude modification. In above, the quantities $u_0(r)$ and $u_M(r)$ are the radial mode functions of the corresponding lasers (now in the detection fiber) and the integers $l_0$ and $l_M$ are the corresponding angular momentum numbers. $\beta_0$ and $\beta_M$ are the propagation constants of the modes along the fiber $z$ axis  and the quantities $\omega_0$ and $\omega_M$ are the angular frequencies. 

We next solve for the propagation of the probe laser field in the presence of the axion field $\phi$ of Eq.~(5). The detailed procedure is outlined in the Appendix~A. Briefly, we first reduce Maxwell's equations for the probe laser beam [Eq.~(2)] into a single wave equation. We then solve this wave equation in the detection fiber, while taking into account the interaction of the probe laser with the axion field and the mixing laser beam. We make several reasonable simplifications and three key assumptions: (i) {\it Energy conservation:} we take the frequencies of the interacting laser beams in the four-wave mixing process to form a closed loop: $\omega_P- \omega_S + \omega_M = \omega_0$. (ii) {\it Angular momentum conservation:} we take the angular momentum numbers for the four laser beams to form a closed loop as well: $l_P - l_S + l_M = l_0$. (iii) We make the Slowly Varying Envelope Approximation (SVEA) for the probe laser field, $ \vert \frac{d E_0} { d z } \vert << \beta_0 E_0$. Under these assumptions and simplifications, the propagation of the probe laser electric field amplitude, $E_0(z)$, can be reduced to a single differential equation:
\begin{eqnarray}
 2 i \beta_0 \frac{ d E_0} {d z} = \frac{g_{a \gamma \gamma}^2} {c} \frac{1}{\kappa_{axion}^2}  (\omega_P - \omega_S) \omega_0  \left(\frac{1}{2 \mu_0} B_M B_S^* \right) E_P \exp{[i (\beta_P - \beta_S + \beta_M - \beta_0) z]} \quad . 
\end{eqnarray}

\noindent Here, the quantity $\beta_P - \beta_S + \beta_M - \beta_0$ is the phase-mismatch of the four-wave mixing interaction. We next define $\Delta k _{FWM} \equiv \beta_P - \beta_S + \beta_M - \beta_0$ to simplify the notation in Eq.~(9) further. We also define $n_{eff} \equiv \frac{\beta_0} {\omega_0/c}$, which can be thought as the effective refractive index for the probe laser mode as it is propagating through the detection fiber. With these definitions, the differential equation that describes the propagation of the probe laser reduces to: 
\begin{eqnarray}
\frac{d E_0}{d z} =  i \xi E_P  \exp{(i \Delta k_{FWM} z)} \quad , 
\end{eqnarray} 

\noindent where the quantity $\xi$ essentially summarizes the whole interaction and is given by:
\begin{eqnarray}
\xi= g_{a \gamma \gamma}^2 \frac{1}{n_{eff}} \frac{1}{c}  \frac{1}{\kappa_{axion}^2} \left( \frac{1}{2 \mu_0} B_M B_S^* \right) (\omega_P - \omega_S)    \quad . 
\end{eqnarray}

We next focus on the ideal phase-matched case where we assume that the mode propagation constants can be adjusted such that $\Delta k_{FWM} \rightarrow 0$ (the case of finite $\Delta k_{FWM}$ is discussed in Appendix~B). In this limit, the probe propagation equation has a particularly simple form and can immediately be solved:
\begin{eqnarray}
\frac{d E_0}{d z} & = &  i \xi E_P   \nonumber \\
\Rightarrow E_0(L) & = & E_0(0) + i  \xi E_P L  \quad . 
\end{eqnarray} 

\noindent Here, $L$ is the total length of each fiber, which can be viewed as the interaction length. Equation~(12) describes the change in the electric field of the probe laser beam due to the axion interaction. Using this change in the electric field, we can also find the corresponding fractional change in the intensity of the probe laser beam:
\begin{eqnarray}
\frac{I(L)}{I(0)} = 1 + 2 \xi L \frac{\Im{(E_P)}}{E_0(0)} \quad .  
\end{eqnarray}

\noindent Here, for simplicity, we have taken the quantities $\xi$ and $E_0(0)$ to be real. In Eq.~(13), the symbol $\Im$ stands taking the imaginary part of the quantity inside the brackets. Note that by changing the phase of the pump beam, $E_P$, we can change the sign of the quantity $\Im{(E_P)}$, and thereby control whether the probe beam will experience absorption or amplification due to axion-mediated four-wave mixing interaction. 

\section{Bounds on the coupling constant}

We next discuss the bounds on the axion coupling constant that such an experiment can place, given specific experimental conditions. The sensitivity of such an experiment will critically depend on to what precision we can measure the change in the intensity of the probe laser beam. For this purpose, to first order, we assume shot-noise limited detection, with fluctuations of order $\sqrt{N_{photon}}$, for a total detected number of probe photons of $N_{photon}$. Under these assumptions and simplifications, given an experimental set of parameters, the bound on the square of axion-photon coupling constant from Eq.~(13) is:
\begin{eqnarray}
g_{a \gamma \gamma}^2 = \frac{ n_{eff} \kappa_{axion} }{2 L  \left( \frac{1}{2 \mu_0} B_M B_S^* \right) \sqrt{N_{photon}} \sqrt{\frac{P_P}{P_0}}} \quad .
\end{eqnarray}

\noindent Here, the quantity  $\frac{1}{2 \mu_0} B_M B_S^*$ can be thought as the magnetic energy density, $P_P$ is the optical power in the pump laser and $P_0$ is the optical power in the probe beam. The details of deriving Eq.~(14) is given in Appendix~B.  The dependence of the coupling constant on the probe refractive index, $n_{eff}$, is in the spirit of enhanced nonlinearities in near-zero index materials \cite{caspani,liberal}.  

 We next evaluate the bounds on the coupling constant for four envisioned phases of the experiment. The parameters that are used in these four phases are given below, in Table-I. The envisioned parameters for the lasers are well within the current the state of the art of high-power fiber lasers \cite{laser_review}. We envision that the parameters that are used in the first two phases of the experiment ({\it phase-1} and {\it phase-2}) can be achieved in a few year time-scale, while the last two phases of the experiment ({\it phase-3} and {\it phase-4}) can be performed within the next 5 to 10 years. There will likely be two main systematics in the experiment: (1) Optical four-wave mixing interaction due to possible residual leakage of the pump and Stokes laser beams to the detection fiber. (2) The change in the intensity of the probe laser beam due to off-resonant Raman scattering in the detection fiber.  To put the parameters that are listed in Table~1 into perspective, the ALPS~II project, which is the next generation LSW experiment currently at construction at DESY, will use a cavity with a finesse of $F=4 \times 10^4$, and a cavity length of $L=88$~m, with a quite stringent DC magnetic-field of 5.3~T to be maintained within the whole length using dipolar magnets \cite{review}. The first results from the ALPS~II project is expected in 2022.

\vspace{0.5cm}

\begin{table}

\begin{tabular}{ || c | c | c | c | c ||}
\hline
Parameter & Phase-1 & Phase-2 & Phase-3 & Phase-4 \\
\hline
Length of the fiber ($L$) & $1$~km & $10$~km & $100$~km & $1000$~km \\ 
\hline
Optical power in probe ($P_0$) & $1$~mW & $10$~mW & $100$~mW & $1$~W  \\
\hline
Integration time ($T$) & $100$~s & $10^3$~s & $10^4$~s & $10^6$~s \\
\hline
Optical power in pump ($P_P$) & $1$~W & $10$~W & $100$~W & $10$~kW \\
\hline
Optical intensity of Stokes  ($I_S$) & $0.1$~GW/cm$^2$ & $1$~GW/cm$^2$ &  $10$~GW/cm$^2$ &  $100$~GW/cm$^2$ \\
\hline 
Optical intensity of mixing ($I_M$) & $0.1$~GW/cm$^2$ & $1$~GW/cm$^2$ &  $10$~GW/cm$^2$ &  $100$~GW/cm$^2$ \\
\hline
The refractive index for probe ($n_{eff}$) & $1$ & $10^{-1}$ & $10^{-2}$ & $10^{-4}$ \\
\hline
\end{tabular}

\caption{The set of parameters that are used for the four envisoned phases of the experiment.}

\end{table}

\vspace{0.5cm}

The calculated bounds for the axion-photon coupling for these four envisioned phases of the experiment are shown in Fig.~4. For comparison the predicted sensitivities of the next generation LSW experiment (ALPS~II), as well  the next generation solar helioscope (IAXO) are also plotted (dashed brown lines) \cite{review}. Solar helioscope IAXO is proposed to utilize 25-m-long, 5.2-m-diameter toroid magnet assembly. The envisioned {\it phase-4} experiment of our scheme (solid black line) is quite competitive with planned ALPS~II and IAXO \cite{review}. 

\begin{figure}[tbh]
\vspace{0cm}
\begin{center}
\includegraphics[width=19cm]{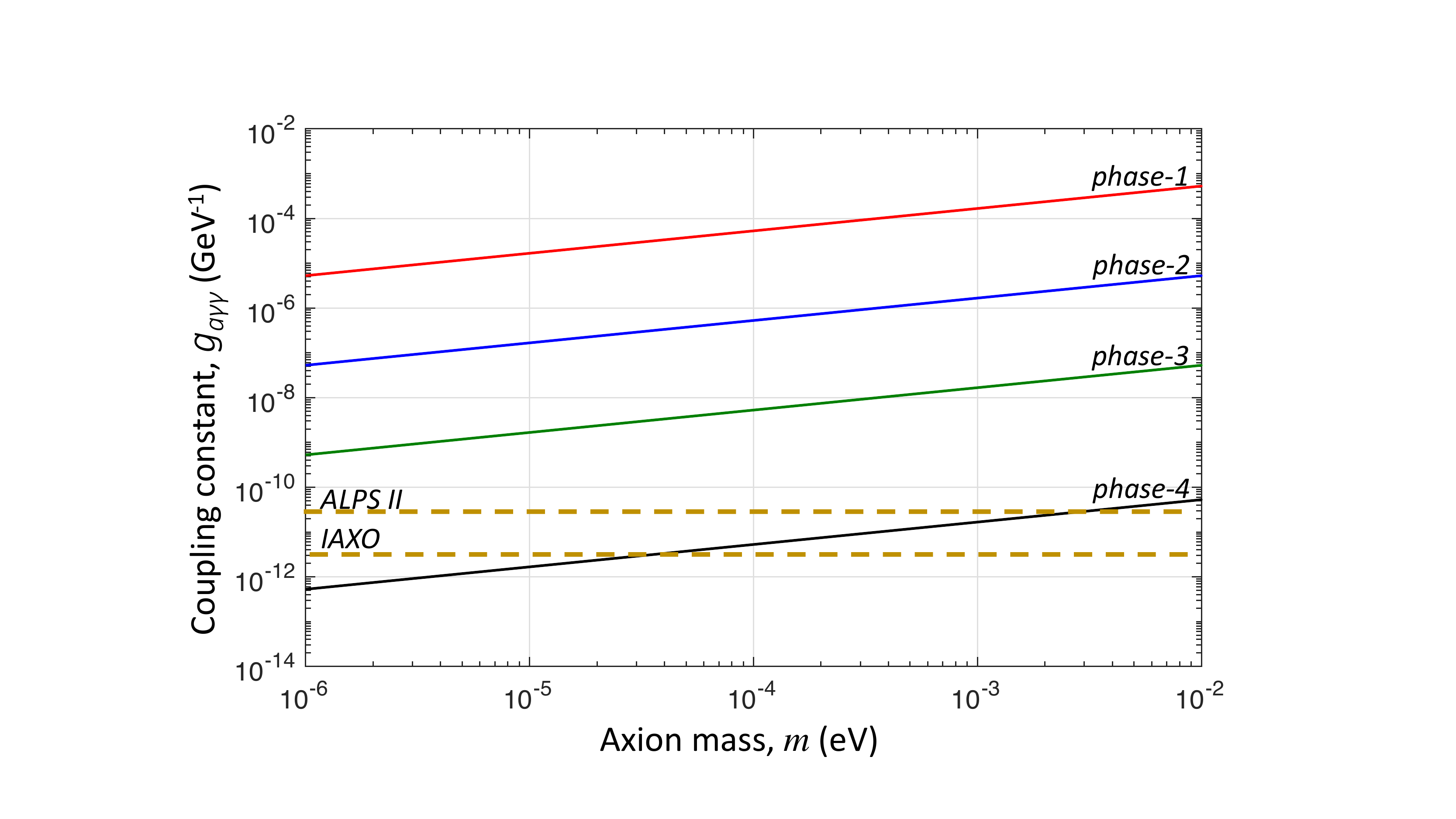}
\vspace{-1cm} 
\caption{\label{eit_scheme} \small  The calculated detection sensitivity for the axion-photon coupling constant for four different phases of the experiment. The {\it phase-4} experiment is quite competitive with several planned experiments, such as the next generation LSW experiment (ALPS~II) and the next generation solar helioscope (IAXO).   }
\end{center}
\vspace{-0cm}
\end{figure}

\section{Conclusions and future directions}

In summary, we have suggested a new approach for generating and detecting axions using lasers in a guided wave geometry. As we mentioned above, there are important advantages of our scheme including the ability to scan a wide mass-range for the axion, and the long interaction length. Furthermore, our scheme does not rely on interaction of the axions with a Tesla level DC magnetic field. The calculations of Fig.~4 predict that our scheme can achieve sensitivities quite competitive to several planned experiments in the near future. While the parameters that are used for the {\it phase-4} calculation in Fig.~4 are stringent, they are within the state of the art. 

We believe there are several promising avenues for future research which may result in significant enhancement of the detection sensitivity in our scheme: (1) The interaction length can further be enhanced by integrating a high-finesse cavity with the detection and generation fibers. Using a distributed Bragg reflector at each end of the fiber (which may be achieved by modulating the refractive index over a section near the beginning and the end each fiber), the lasers can be made to bounce back and forth along the fiber, thereby significantly increasing the interaction length. For this case, the sensitivity bounds for the axion-photon coupling constant $g_{a \gamma \gamma}$ would reduce by a factor $\sqrt{F}$, where $F$ is the cavity finesse. (2) For the calculation of Fig.~4, we assumed shot noise limited detection for the probe laser beam. If squeezed light for the probe laser beam is used, the fluctuations in either the phase or the intensity of the probe laser can be reduced substantially below the shot-noise limit, thereby resulting in improved sensitivity for axion detection. (3) It may also be possible to utilize metamaterials in other ways to increase the detection sensitivity. For example, waveguides with a high effective magnetic susceptibility may be used to increase the magnetic field values for the Stokes and Mixing lasers [i.e., it may be possible to enhance the values of the quantities $B_S$ and $B_M$ in Eq.~(14)] \cite{caspani,liberal}.

We thank Ben Lemberger for many helpful discussions. This work was supported by the University of Wisconsin-Madison through the Vilas Associates Award.

\newpage

\section{Appendix A: derivation of the propagation equation for the probe laser}

In this section, we will discuss the derivation of the Slowly Varying Envelope Approximation (SVEA) propagation equation of Eq.~(9 for the probe laser beam. We first follow the usual procedure and combine Maxwell's equations of Eq.~(2) into a single wave equation for the probe laser Electric field:
\begin{eqnarray}
\nabla \times \vec{E} & = & -\frac{ \partial \vec{B}}{\partial t}  \quad , \nonumber \\
\Rightarrow \nabla \times \nabla \times \vec{E}  & = & -\frac{ \partial }{\partial t} \nabla \times \vec{B}  \quad , \nonumber \\
\Rightarrow \nabla \left( \nabla \cdot \vec{E}  \right)  - \nabla^2 \vec{E} & = &  -\frac{ \partial }{\partial t} \left[ \frac{1}{c^2} \frac{ \partial \vec{E}}{\partial t}  + \mu_0 \vec{J} + \frac{g_{a \gamma \gamma}}{c} \left( \frac{\partial \phi}{\partial t} \vec{B} + \nabla \phi \times \vec{E} \right) \right]  \quad ,  \nonumber \\
\Rightarrow \nabla^2 \vec{E} & = & \frac{n^2}{c^2} \frac{ \partial^2 \vec{E}}{\partial t^2} + \frac{g_{a \gamma \gamma}}{c} \left[ \frac{\partial^2 \phi}{\partial t^2} \vec{B}+ \frac{\partial \phi}{\partial t} \frac{ \partial \vec{B}}{\partial t} + \frac{\partial} {\partial t} \left( \nabla \phi \times \vec{E} \right) \right] \quad .
\end{eqnarray}

\noindent Here, in the last line, we have summarized the effect of the current density of the material, $\vec{J}$, into a refractive index for the probe laser using the relative permittivity $\epsilon_r$ and the well-known relations: $\vec{J} = \epsilon_r \epsilon_0 \frac{ \partial \vec{E} } {\partial t}$ and  $ n^2 = 1 + \epsilon_r$:
\begin{eqnarray}
\frac{1}{c^2} \frac{ \partial \vec{E}}{\partial t}  + \mu_0 \vec{J} = \frac{1}{c^2} \frac{ \partial \vec{E}}{\partial t}  + \mu_0 \epsilon_r \epsilon_0 \frac{ \partial \vec{E}} {\partial t} = \frac{(1 + \epsilon_r)}{c^2} \frac{ \partial \vec{E} } {\partial t} = \frac{n^2}{c^2} \frac{ \partial \vec{E} } {\partial t} \quad . 
\end{eqnarray}

\noindent The refractive index is a function of the radial coordinate, $n(r)$, which is required to have a confined mode for the probe laser beam in the detection fiber. There is another step used in the last line of Eq.~(15), which is ignoring the divergence of the probe electric field: $\nabla \cdot \vec{E}  =  - g_{a \gamma \gamma} c \nabla \phi \cdot \vec{B} \approx 0$. This step is not strictly necessary, but greatly simplifies the algebra of what follows. We, therefore, take the parameters of the system such that the divergence of the probe electric field can be ignored. Specifically, using $\nabla \phi \cdot \vec{B} = \frac{\partial \phi} {\partial r} B_r + \frac{\partial \phi} {\partial \varphi} B_{\varphi}$, we make two reasonable simplifications: (1) We take axion field to vary sufficiently slowly as a function of the radial coordinate along the detection fiber such that $\frac{\partial \phi} {\partial r} \approx 0$. (2) Noting Eq.~(5) for the axion field solution, we consider pump and Stokes driving laser modes such that $l_P = l_S$, and therefore $\frac{\partial \phi} {\partial \varphi} = 0$. 

We next use the axion field solution of Eq.~(5) and the forms for the probe and mixing lasers of Eq.~(7), to reduce the wave equation for the probe electric field to:
\begin{eqnarray}
&  & \left[ \frac{d^2 u_0} {d r^2} + \frac{1}{r} \frac{d u_0}{d r} - \frac{l_0^2 }{r^2}u_0 \right] E_0 \exp{(i l_0 \varphi )} \exp{ [i ( \beta_0 z - \omega_0 t)] } \quad \nonumber \\
& + & \left[ \frac{ d^2 E_0 } {d z^2} - 2 i \beta_0 \frac{d E_0} {d z} - \beta_0^2 E_0 \right] u_0 \exp{(i l_0 \varphi )} \exp{ [i ( \beta_0 z - \omega_0 t)] } \quad \nonumber \\ 
& = & - \frac{n^2}{ c^2} \omega^2 u_0 \exp{(i l_0 \varphi )} \exp{ [i ( \beta_0 z - \omega_0 t)] } \quad \nonumber \\ 
& - & \frac{ g_{a \gamma \gamma}} {c} B_M (\omega_P - \omega_S)^2 u_{\phi} u_M  \exp{[i (l_P-l_S+l_M) \varphi ]} \exp{ [i ( \beta_P  - \beta_S + \beta_M)z -i (\omega_P -\omega_S +\omega_M) t]} \quad \nonumber \\ 
& - & \frac{ g_{a \gamma \gamma}} {c} B_M (\omega_P - \omega_S) \omega_M u_{\phi} u_M  \exp{[i (l_P-l_S+l_M) \varphi ]} \exp{ [i ( \beta_P  - \beta_S + \beta_M)z -i (\omega_P -\omega_S +\omega_M) t]}  \quad . \nonumber \\
\end{eqnarray}

\noindent We next simplify Eq.~(17) using the following assumptions: (i) {\it Energy conservation:} we take the frequencies of the interacting laser beams in the four-wave mixing process to form a closed loop: $\omega_P- \omega_S + \omega_M = \omega_0$. (ii) {\it Angular momentum conservation:} we take the angular momentum numbers for the four laser beams to form a closed loop as well: $l_P - l_S + l_M = l_0$. (iii) We make the Slowly Varying Envelope Approximation (SVEA) for the probe laser field, $ \vert \frac{d E_0} { d z } \vert << \beta_0 E_0$. We also note that, without the axion interaction, the probe wave is a mode of the waveguide, i.e., the radial profile for the probe laser beam satisfies:
\begin{eqnarray} 
\frac{d^2 u_0} {d r^2} + \frac{1}{r} \frac{d u_0}{d r} + \left( \frac{n^2}{c^2} \omega_0^2 - \beta_0^2 -\frac{l_0^2 }{r^2} \right) u_0 = 0 \quad . 
\end{eqnarray}

\noindent With these assumptions and simplifications, Eq. (17) reduces to:
\begin{eqnarray}
- 2 i \beta_0 \frac{d E_0} {d z} u_0 = -\frac{g_{a \gamma \gamma}}{c} (\omega_P - \omega_S) \omega_0 B_M u_{\phi} u_M \exp{ [i ( \beta_P  - \beta_S + \beta_M - \beta_0)z ]]} \quad . 
\end{eqnarray}

\noindent To simplify Eq.~(19) further, we take the probe and the mixing beam radial profiles to be similar so that $u_0 (r) \approx u_M(r)$. We will also take the radial variation of the axion field over the detection fiber to be negligible, and assume $u_{\phi}(r) \approx$~constant. Because Eq.~(7) is in normalized units, up to a numerical factor, the value of the axion field is $u_{\phi} (r) \sim \frac{1}{\kappa_{axion}^2} \frac{g_{a \gamma \gamma}}{\mu_0 c} E_P B_S^*$. Using this value for the axion field radial value, $u_{\phi}$ in Eq.~(19) then gives the probe propagation equation of Eq.~(9). 

\section{Appendix B: derivation of the bound on the coupling constant}

Assuming shot noise limited detection, the experiment will be sensitive to the intensity change of the probe laser beam at the level of $1/\sqrt{N_{photon}}$. The predicted intensity change of the probe laser beam due to axion interaction is given by the second term in the right hand side of Eq.~(13). Equating this change to $1/\sqrt{N_{photon}}$, we get:
\begin{eqnarray}
\vert 2 \xi L \frac{\Im{(E_p)}}{E_0(0)} \vert = \frac{1}{\sqrt{N_{photon}}} \quad . 
\end{eqnarray}

We note that the ratio of the field amplitude of the pump to the probe laser beam can be expressed as the square-root of the ratio of the corresponding optical powers, assuming similar mode sizes:
\begin{eqnarray}
\frac{\Im{(E_p)}}{E_0(0)} \approx \sqrt{\frac{P_P}{P_0}} \quad .
\end{eqnarray}

We obtain Eq.~(14) by using Eq.~(21) and the expression for the quantity $\xi$ from Eq.~(11) of above, in Eq.~(20). 

As we discussed above, the bound for the axion-photon coupling constant of Eq.~(14) assumes the ideal case of perfect phase-matching for the four-wave mixing interaction, $\Delta k_{FWM}=0$. For finite $\Delta k_{FWM}$, the solution of Eq.~(10) is:
\begin{eqnarray}
E_0(L) = E_0(0) + i \xi E_P \frac{\sin{(\frac{\Delta k_{FWM}L }{2})}}{\frac{\Delta k_{FWM} L}{2}} \exp{(i \Delta k_{FWM} L /2)} L \quad . 
\end{eqnarray}

The extra $\frac{\sin{(\frac{\Delta k_{FWM}L }{2})}}{\frac{\Delta k_{FWM} L}{2}}$ factor for finite $\Delta k_{FWM}$ can be carried out through the rest of the formalism, which would then modify the bound on the axion-photon coupling constant that is given in Eq.~(14). To first order, this would reduce the sensitivity of the scheme by a factor of, $\Delta k_{FWM} L/2$.

\end{document}